\documentstyle[12pt,epsf,epsfig,wrapfig]{article}
\newcommand{\la}{$\Lambda$}
\newcommand{\al}{$\bar{\Lambda}$}

\begin{document}

\begin{center}
{\bfseries COMPARISON OF LONGITUDINAL POLARIZATION OF $\Lambda$ AND
$\bar\Lambda$ IN DEEP-INELASTIC SCATTERING  AT COMPASS}

\vskip 5mm M.G.Sapozhnikov\\ on behalf of the COMPASS Collaboration

\vskip 5mm {\small  {\it Joint Institute for Nuclear Research,
Laboratory of Particle Physics, Dubna
}\\

{\it E-mail: sapozh@sunse.jinr.ru }}
\end{center}

\vskip 5mm
\begin{abstract}
The longitudinal polarization of $\Lambda$ and $\bar \Lambda$
produced in deep-inelastic scattering of 160 GeV/{\it c} polarized
muons is studied in the COMPASS experiment. Preliminary results on
$x-$ and $y-$ dependence of the longitudinal polarization of \la~
and \al~ from data collected during the 2003 run  are presented.
\end{abstract}

\vskip 8mm

The study of longitudinal polarization of $\Lambda$ and \al~
hyperons in the deep-inelastic scattering (DIS) is important for
understanding  fundamental properties of the nucleon. Comparing the
longitudinal polarizations of $\Lambda$ and \al~ in DIS one could
test if  strange $s(x)$ and antistrange $\bar{s}(x)$ quark
distributions are equal and, in principle, it would be possible to
obtain information about polarization of the strange quarks in the
nucleon \cite{Ell.95}-\cite{Ell.00}. The useful information about
the spin structure of \la~ could be obtained
\cite{BJ}-\cite{Yan.00}.

The polarized nucleon intrinsic strangeness model
\cite{Ell.95,Ell.00}
 predicts negative longitudinal polarization
of $\Lambda$ hyperons produced in the target fragmentation region
\cite{Ell.96,Ell.02}. The main assumption of the model is the
negative polarization of the strange quarks and antiquarks in the
nucleon. This assumption was inspired by the results of EMC
\cite{EMC.88} and subsequent experiments \cite{SMC.97}- \cite{E155}
on inclusive deep-inelastic scattering which gave indication that
the $s\bar{s}$ pairs in the nucleon are negatively polarized with
respect to the nucleon spin:
\begin{equation}
\Delta s \equiv \int\limits_{0}^{1} dx[ s_{\uparrow}(x) -
s_{\downarrow}(x) + \bar s_{\uparrow}(x)- \bar s_{\downarrow}(x)] =
- 0.10\pm0.02.
\end{equation}

The prediction of the model was confirmed in the neutrino DIS
experiments, where large and negative longitudinal polarization
$P_{\Lambda}$ of the $\Lambda$ hyperons at the target fragmentation
region was found. The best statistics was obtained in the NOMAD
experiment \cite{NOMAD}, where $P_{\Lambda}=-0.21\pm0.04\pm0.02$ was
measured at $x_F<0$.

Negative polarization of the strange sea was also found in the
recent QCD analysis of the available world data \cite{Lea.05}: $
\Delta s= -0.049 \pm 0.013$. However the question about polarization
of the nucleon strange quarks can not be considered to be solved.
Thus the HERMES collaboration, after analysis of the semi-inclusive
DIS channels in the LO approximation,
 found that $\Delta s =0.028\pm0.033\pm0.009$
\cite{HERMES.04}, i.e. consistent with zero within the errors (for
the discussion of the HERMES result, see
\cite{Kot.04},\cite{Lea.03}). Therefore, the possibility to obtain
an information on  $\Delta s$ from  data on longitudinal
polarization of \la~ and \al~ is  quite important.

The  $\Lambda$ production in the current fragmentation region
$x_F>0$ is believed to be dominated by the struck quark
fragmentation. Then the longitudinal $\Lambda$ polarization in the
parton model is determined as follows \cite{Kot.98}:

\begin{equation}
 P_{\Lambda}=\frac{\sum_{q} e^2_q [P_b D(y) q(x) + P_T \Delta q(x)]
\Delta D^{\Lambda}_q(z)}{\sum_{q} e^2_q [ q(x) + P_b D(y) P_T \Delta
q(x)] D^{\Lambda}_q(z)} \label{plam}
\end{equation}

Here $e_q$ is the quark charge, $P_b$ and $P_T$ are the polarization
of the beam and target, $q(x)$ and $\Delta q(x)$ are unpolarized and
polarized quark distribution functions, $D^{\Lambda}_q(z)$ and
$\Delta D^{\Lambda}_q(z)$ are unpolarized and polarized
fragmentation functions,  $D(y)$ is the longitudinal depolarization
factor of the virtual photon with respect to the initial state
lepton, $y$ is the fraction of the lepton energy carried out by the
virtual photon $y=\nu/E$ ($\nu=E-E'$, $E$ and $E'$ are lepton
energies in the initial and final states), $z$ is the fractional
hadron energy, $z=E_h/\nu$, $x$ is the Bjorken scaling variable.

The expression (\ref{plam}) helps to understand salient features of
the spin transfer to \la. For the scattering on the unpolarized
target $P_T=0$, assuming that only one quark flavor dominates, it
may be expected that the polarization will be proportional to the
$D(y)$. The depolarization factor $D(y)$ increases with $y$,
therefore one should expect that the $P_{\Lambda}$ will also grow up
with $y$.

From (\ref{plam})  it is clear that the polarizations of \la~and
\al~ should be the same if we are in the region where
$q(x)=\bar{q}(x)$ and $\Delta q(x)= \Delta \bar{q}(x)$. Again this
statement is valid if the \la~ has been produced mainly due to quark
fragmentation and if the contribution from the diquark fragmentation
is small.

Essential ingredient of the (\ref{plam}) is the polarized
fragmentation function $\Delta D^{\Lambda}_q(z)$. There are
different approaches to the choice of $\Delta D^{\Lambda}_q(z)$
\cite{BJ}-\cite{Yan.00}. All logically possible values of the
\la~polarization (positive, negative and zero) have been predicted.

 In the naive quark model (NQM) the spin of $\Lambda$ is carried by
the $s$ quark and the spin transfer from the $u$ and $d$ quarks to
$\Lambda$ is equal to zero. It means that if we are in the region
where $\Lambda$ is produced in the fragmentation of $u$ and $d$
quarks, one may expect that $P_{\Lambda} \sim 0$.

The authors of \cite{BJ}, using $SU(3)_f$ symmetry and experimental
data for the spin-dependent quark distributions in the proton,
predict that the contributions of $u$ and $d$ quarks to the
$\Lambda$ spin are negative and substantial, at the level of 20\%
for each light quark. In this model the fragmentation of the
dominant $u$ quark will lead to negative spin transfer to \la.

Positive spin transfer to \la~ has been predicted in the advanced
NQM model \cite{BGH}, where indirect production of \la~from decays
of heavy hyperons (such as $\Sigma, \Sigma(1385), \Xi$) were also
considered. It was assumed that, as in NQM, the $s$ quarks is
responsible for the spin transfer to \la ~ produced directly. The
spin transfer from $u$- and $d$-quarks is also possible due to
production of polarized heavy hyperons, which decayed into \la~,
having inherited a part of its polarization.

Positive spin transfer from $u$- and $d$-quarks was also predicted
in the framework of $SU(6)$ based quark-diquark model
\cite{Ma.00},\cite{Yan.00}. They found  large and positive
polarization of the $u$ and $d$ quarks in the $\Lambda$ at large
$x$.

The assumption about dominance of the quark fragmentation processes
for the \la~ production was questioned in \cite{Ell.02}. It has been
shown that the energies of the current experiments are not large
enough and even at the COMPASS energy of 160 GeV most of $\Lambda$,
even in the $x_F>0$ region, are produced from the diquark
fragmentation.

 We have studied $\Lambda$ and $\bar \Lambda$  production by
polarized $\mu^+$ of 160 GeV on a polarized $^{6}$LiD target of the
 spectrometer constructed in the framework of COMPASS experiment
(NA58) at CERN. A detailed description of the COMPASS experimental
setup is given elsewhere \cite{COMPASS} (see, also talks of
F.Bradamante and Y.Bedfer at this conference) and only the most
relevant elements for the present analysis will be given below.

The muon beam polarization is $P_b=-0.76\pm0.04$. The polarized
target consists of two oppositively polarized cells, 60 cm long and
3 cm in diameter, separated by 10 cm. The cells are located on the
axis of a superconducting solenoid magnet providing a field of 2.5 T
along the beam direction, and are filled with $^{6}$LiD . This
material is used as a deuteron target and was selected for its high
dilution factor of about 40\%, which accounts for the fact that only
a fraction of the target nucleons are polarizable. Typical
polarization values of 50\% are obtained. The two cells are
polarized in opposite directions by using different microwave
frequencies so that data with both spin orientations are recorded
simultaneously.
 For this analysis the data are averaged over the target
polarization.

The COMPASS spectrometer has a large and  small angle spectrometers
built around two dipole magnets, in order to allow the
reconstruction of the scattered muon and of the produced hadrons in
broad momentum and angular ranges. Different types of tracking
detectors are used to deal with the rapid variation of the particle
flux density with the distance from the beam. Tracking in the beam
region is performed by scintillating fibers. Up to 20 cm from the
beam we use Micromegas and GEMs. Further away, tracking is carried
out in multiwire proportional chambers and drift chambers.
Large-area trackers, based on straw detectors and large drift
chambers extend the tracking over a surface of up to several square
meters. Muons are identified by dedicated trackers placed downstream
of hadron absorbers. Hadron/muon separation is strengthened by two
large iron-scintillator sampling calorimeters, installed upstream of
the hadron absorbers and shielded to avoid electromagnetic
contamination. The particle identification provided by the ring
imaging Cherenkov detector is not used in the present analysis.

The trigger system \cite{trig} provides efficient tagging down to
$Q^2$ = 0.002 (GeV/c)$^2$, by detecting the scattered muon in a set
of hodoscopes placed behind the two dipole magnets. A large enough
energy deposit in the hadronic calorimeters is required to suppress
unwanted triggers generated by halo muons, elastic muon-electron
scattering events, and radiative events.

The COMPASS data taking was going on in 2002-2004. The preliminary
analysis of the 2002 run was given in \cite{Sap.04},\cite{Alex.04}.
Here we presented the first results of the analysis of data
collected during the 2003 run. The data sample comprises about
$8.7\cdot10^7$ DIS events with $Q^{2}
>1~$ (GeV/{\it c})$^{2}$.

The $V^{0}$ events ($V^0 \equiv \Lambda$, $\bar\Lambda$  and
$K^{0}_{S}$)  were selected by requiring a primary vertex with
incoming and outgoing muon tracks and at least two hadron tracks
forming a secondary vertex. The primary vertex has to be inside the
target. The secondary vertex must be downstream the both target
cells. The angle $\theta_{col}$ between the vector of $V^0$ momentum
and the vector between primary and $V^0$ vertices should be
$\theta_{col}<0.01$ rad. Cut on transverse momentum $p_t$ of the
decay products with respect to the direction of $V^{0}$ particle
$p_t
>23$ MeV/c was applied to reject $e^{+}e^{-}$ pairs from the
$\gamma$ conversion.  We select events with momenta of positive and
negative particles greater than 1 GeV/c. The momenta $p_V$ of the
$V^{0}$ particles have to be larger than 10 GeV/c. The  DIS cuts on
$Q^{2}
>1~$ (GeV/{\it c})$^{2}$ and $0.2< y < 0.9$ have been applied.

To construct angular distributions of $V^0$ events, the so called
bin-by-bin method was used.
\begin{wrapfigure}[21]{R}{9cm}
\begin{center}
\mbox{\epsfig{figure=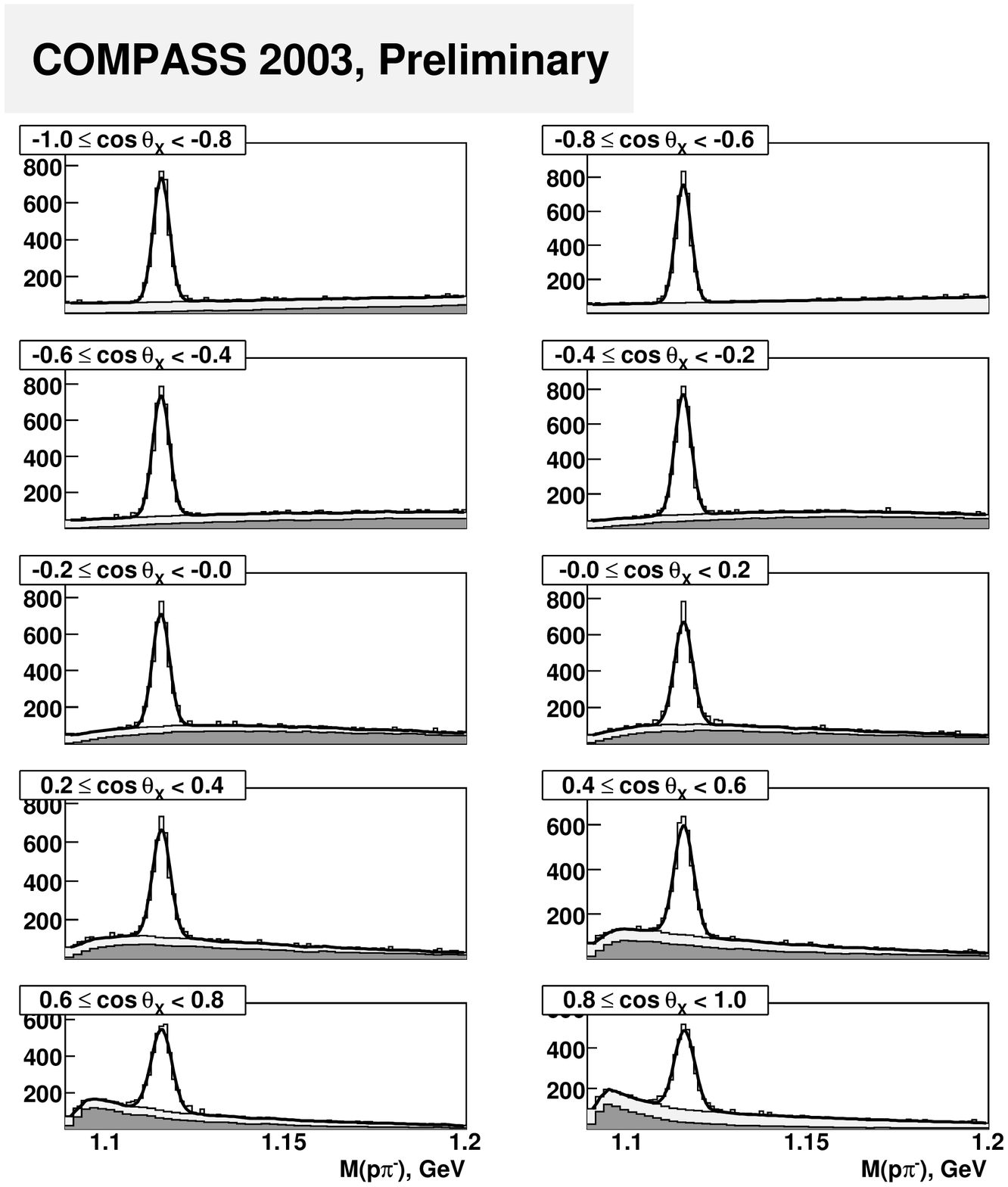,width=9cm,height=9cm}}
\end{center}
{\small{\bf Figure 1.} The invariant mass distribution of $p \pi^-$
  at different $\cos{\theta_X}$ bins.} \label{yourname_fig1}
\end{wrapfigure}
All angular distributions were divided in some bins.  In each  bin
the invariant mass distribution of positive and negative particles
was constructed assuming the $\pi^+ \pi^-$, $p\pi^-$ or $\bar{p}
\pi^+$ hypothesis. The peak of the corresponding $V^0$ particle was
fitted  obtaining  the number of the $V^0$ in the bin. This
procedure allows to construct practically background-free angular
distributions.

An example is shown in Fig.1, where the invariant mass distribution
of $p \pi^-$ at different $\cos{\theta_X}$ bins are shown
($\theta_X$ is the angle between the direction of the decay proton
for \la~  and the direction of the virtual photon in the $V^0$ rest
frame).  A part of the background is coming from kaons, which passed
all selection criteria. We have estimated the kaon background from
the Monte Carlo simulation, it is shown by hatched regions in Fig.1.

 The total data
sample contains about 31000 $\Lambda$ and 18000 $\bar\Lambda$. That
is significantly larger than the data sample of the 2002 run, which
comprises about 9000 $\Lambda$ and 5000 $\bar\Lambda$.

In Fig.2 the experimental distributions on different kinematical
variables for \la~(a) and \al~(b) are shown.

\begin{figure}[htb!]
\begin{center}
\begin{tabular}{cc}
\mbox{\epsfig{figure=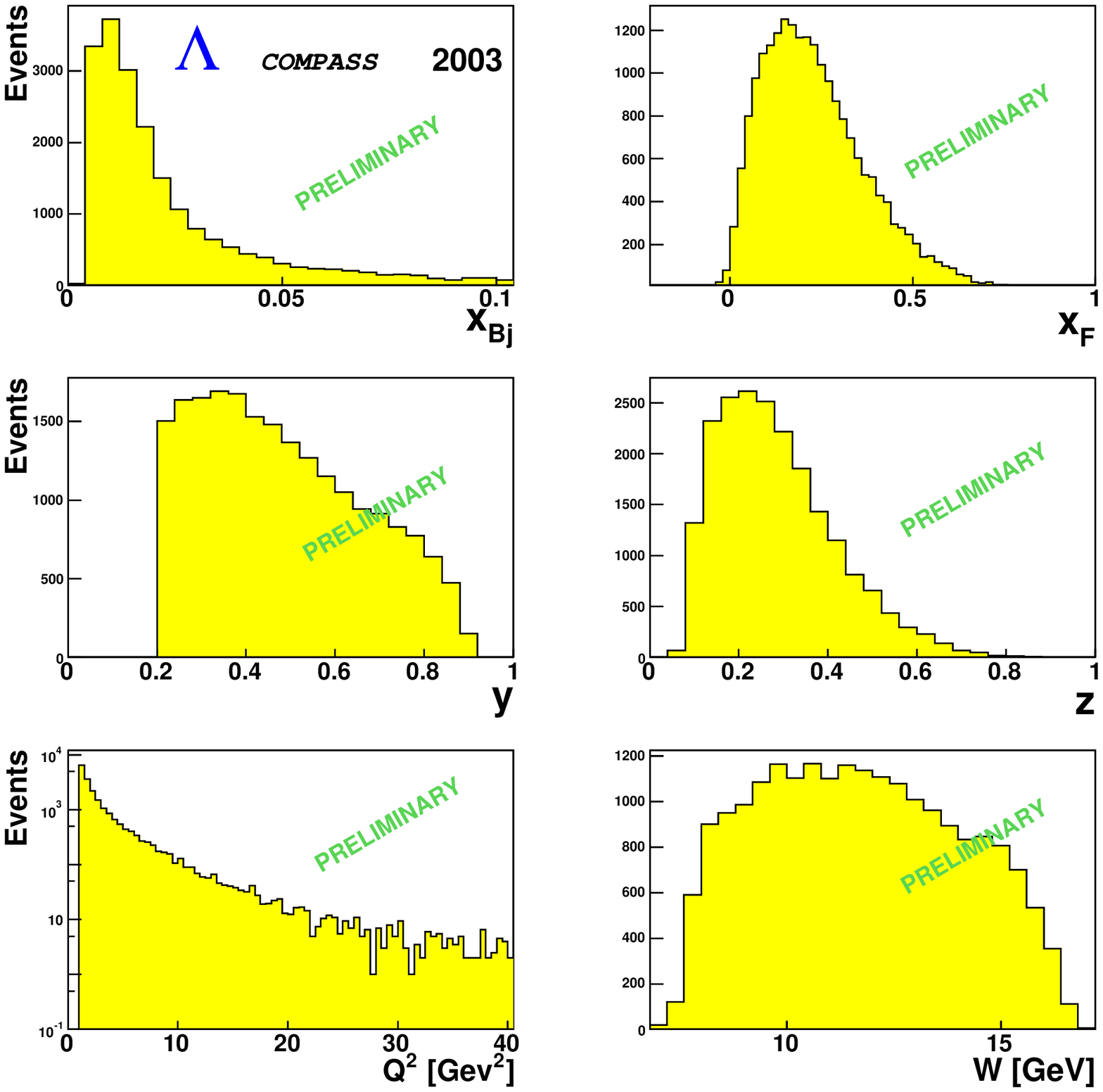,width=6cm,height=7cm}}&
\mbox{\epsfig{figure=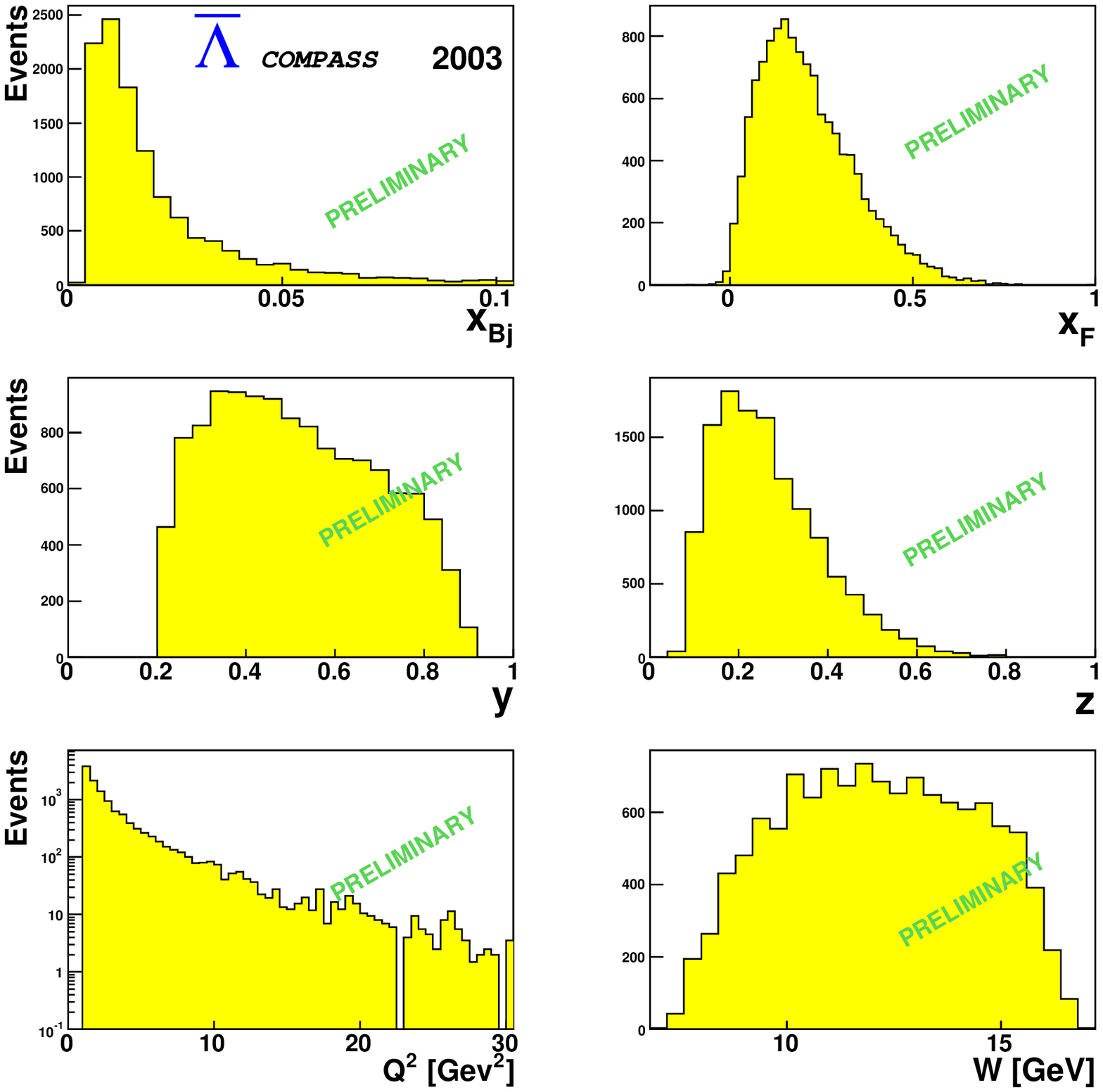,width=6cm,height=7cm}}\\
 {\bf(a)}& {\bf(b)}
\end{tabular}
\end{center}
{\small{\bf Figure 2.} Experimental distributions on $x,  x_F, y, z,
Q^2$ and $W$  for the \la~(a) and \al~(b).} \label{kin}
%{\small{\bf Figure 1b.} Experimental distributions on $Q^2, x_F, y,
%W, p_t$ and $x_{Bj}$  for the \la~hyperons.}
\end{figure}

One can see that our working interval is at a small $x$ region, with
the averaged value $\bar{x}=2.8\cdot10^{-2}$. The COMPASS
spectrometer selects \la~  in the current fragmentation region,
starting from $x_F >-0.1$, with the averaged value $\bar{x_F}=0.23$.
The trigger enriched the data sample with events having high $y$,
the averaged value is $\bar{y}=0.48$. The typical \la~ fractional
energy $z$ is not large $\bar{z}=0.29$. The averaged values of $Q^2$
and $W$ are $\bar{Q^2}=3.55~(GeV/c)^2$ and $\bar{W}=11.7$ GeV,
respectively. The \al~ has very similar distributions of the
kinematical variables.

  The angular distribution  of the decay particles in the $V^0$~
rest frame is

\begin{equation}
w(\theta)=\frac{dN}{d \cos{\theta}}=\frac{N_{tot}}{2}(1+\alpha P
\cos{\theta}), \label{ideal}
\end{equation}
where $N_{tot}$ is the total number of events,
$\alpha=+(-)0.642\pm0.013$ is \la(\al)~decay parameter, $P$ is the
projection of the polarization vector on the direction of the
virtual photon in the $V^0$ rest frame, $\theta$ is the angle
between the direction of the decay proton for \la~ (antiproton - for
\al, positive $\pi$ - for $K^0$) and the direction of the virtual
photon in the $V^0$ rest frame.

Fig.3a shows the measured angular distributions
 for all events of 2002 run for the $K^{0}_{S}$, $\Lambda$ and $\bar\Lambda$ decays, corrected for the acceptance.
The acceptance was determined by the Monte Carlo simulation of
unpolarized $\Lambda(\bar\Lambda)$ decays.

%\begin{figure}[htb]
%\begin{center}
%\epsfig{file=sapozh_fig2.eps,width=10.cm} \caption{\small{\bf Figure
%2.} The angular distributions for $K^{0}_{S}$, $\Lambda$ and
%$\bar\Lambda$ for all events of 2002 run. } \label{sapozh_fig2.eps}
%\end{center}
%\end{figure}

%\begin{figure}[h]
 %\epsfysize=70mm
 %\centerline{
%\epsfbox{sapozh_fig2.eps}}
 %\caption{\small{\bf Figure 2.} The angular distributions for $K^{0}_{S}$,
%$\Lambda$ and $\bar\Lambda$ for all events of 2002 run. }
%\label{sapozh-fig2}
%\end{figure}

\begin{figure}[htb!]
\begin{center}
\begin{tabular}{cc}
\mbox{\epsfig{figure=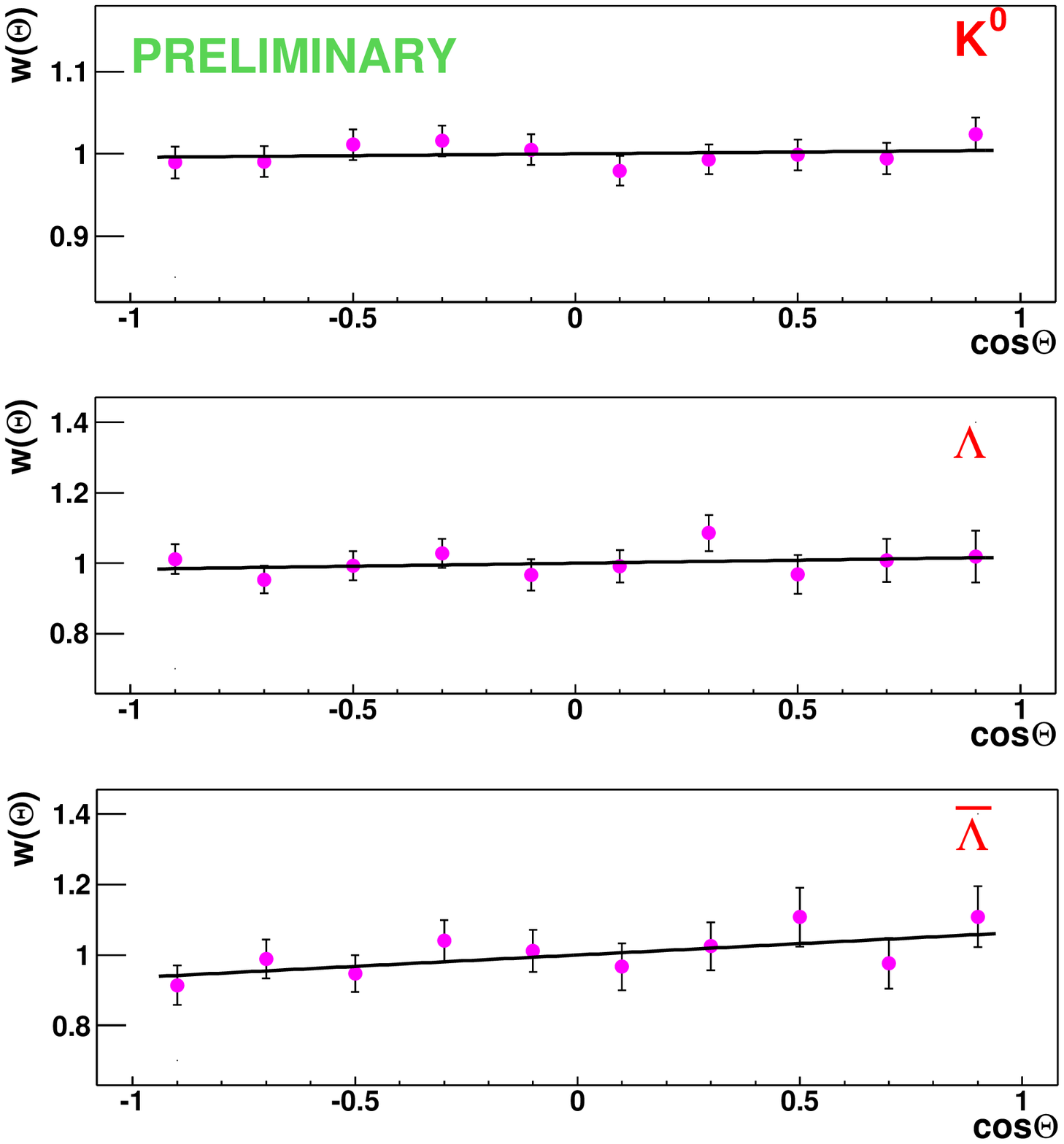,width=7cm,height=9cm}}&
\mbox{\epsfig{figure=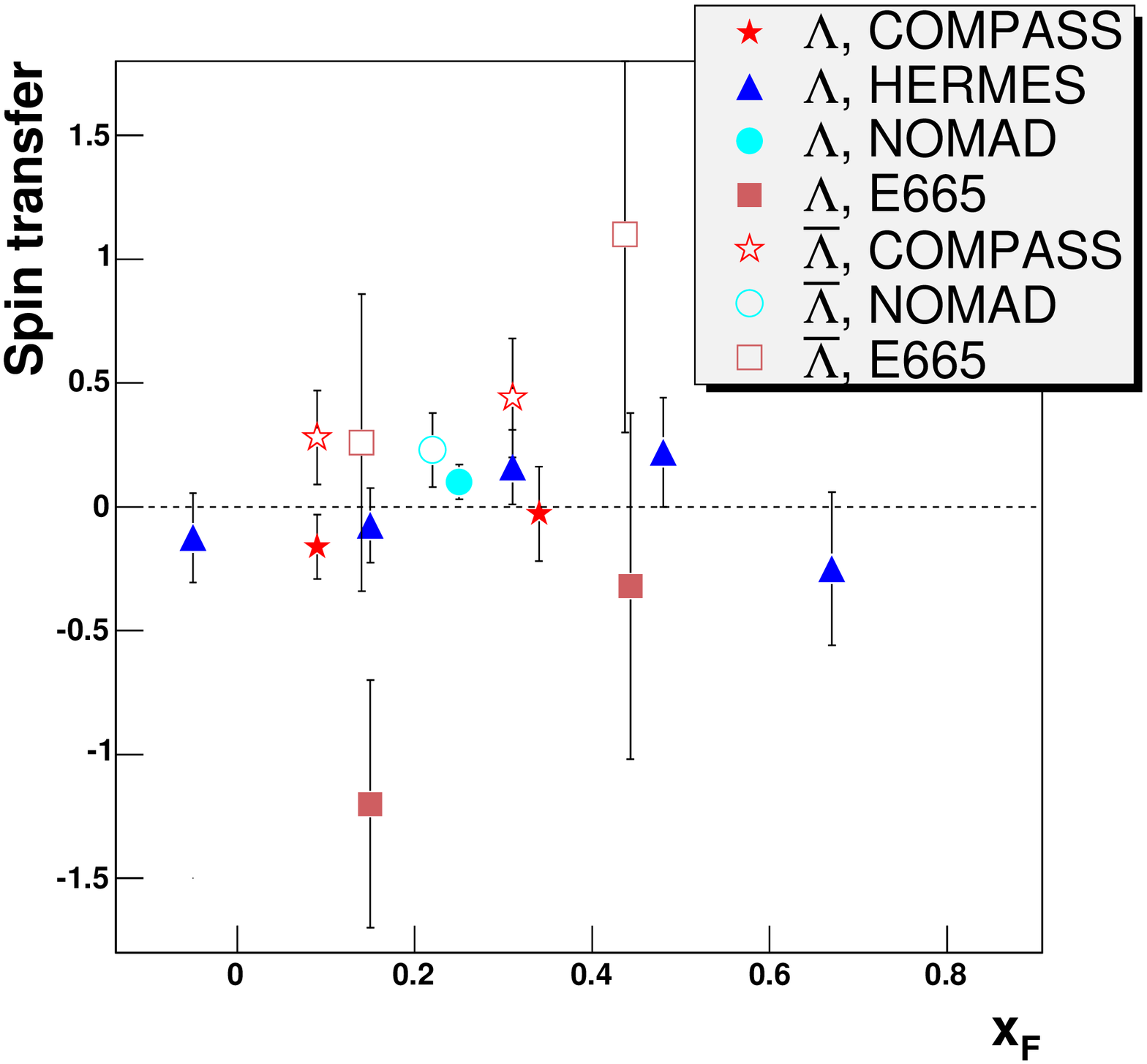,width=7cm,height=9cm}}\\
 {\bf(a)}& {\bf(b)}
\end{tabular}
\end{center}
{\small{\bf Figure 3.} The angular distributions for $K^{0}_{S}$,
$\Lambda$ and $\bar\Lambda$ for all events of 2002 run (a).
Comparison of the spin transfer to \la~ and \al~ hyperons measured
in different DIS experiments \cite{NOMAD},\cite{hermes},\cite{E665}
(b).}
\end{figure}

%\begin{wrapfigure}[21]{R}{6cm}
%\begin{center}
%\mbox{\epsfig{figure=sapozh_fig2.eps,width=6cm,height=6cm}}
%\end{center}
%{\small{\bf Figure 2.} The angular distributions for $K^{0}_{S}$,
%$\Lambda$ and $\bar\Lambda$ for all events of 2002 run.}
%\label{sapozh-fig2}
%\end{wrapfigure}

One could see that the angular distribution for $K^{0}_{S}$ decays
is flat, as expected. The value of the longitudinal polarization is
$P_K=0.007\pm0.017$. The polarization of \la, averaged over all
kinematical variables, is small. The corresponding polarization of
\al~ is small and negative.

The comparison between the COMPASS data of 2002 run on the spin
transfer to \la~ and \al~hyperons with results of other  DIS
experiments is shown in Fig. 3b. The spin transfer $S$ determines
which part of the lepton polarization $P_b$ is transferred to the
hyperon polarization $P$. It is defined as $P=S\cdot P_b\cdot D(y)$,
where $D(y)$ is the virtual photon depolarization factor.

%\begin{wrapfigure}[21]{R}{6cm}
%\begin{center}
%\mbox{\epsfig{figure=sapozh_fig3.eps,width=6cm,height=5cm}}
%\end{center}
%{\small{\bf Figure 3.} Comparison of the spin transfer to \la~(left)
%and \al~(right) hyperons measured in different DIS experiments. }
%\label{lam2}
%\end{wrapfigure}

One can see that there is a reasonable agreement between the COMPASS
and world data. The spin transfer to \la~seems to be small in all
the region of $x_F>0$. The spin transfer to \al~ seems to be
slightly larger but the statistics is not  enough to prove the
existence of this difference.

In Fig. 4a  the y-dependence of the longitudinal polarization is
shown for data of the 2003 run. The data sample  has been divided it
3 bins: $0.2 \leq y < 0.36$, $0.36 \leq y <
    0.55$ and $0.55 \leq y \leq 0.9$. The number of events in each bin
    was approximately the same. The errors on the plot are
    statistical. The systematic error have been evaluated to be
    not larger than 5\% for each data point.

\begin{figure}[htb!]
\begin{center}
\begin{tabular}{cc}
\mbox{\epsfig{figure=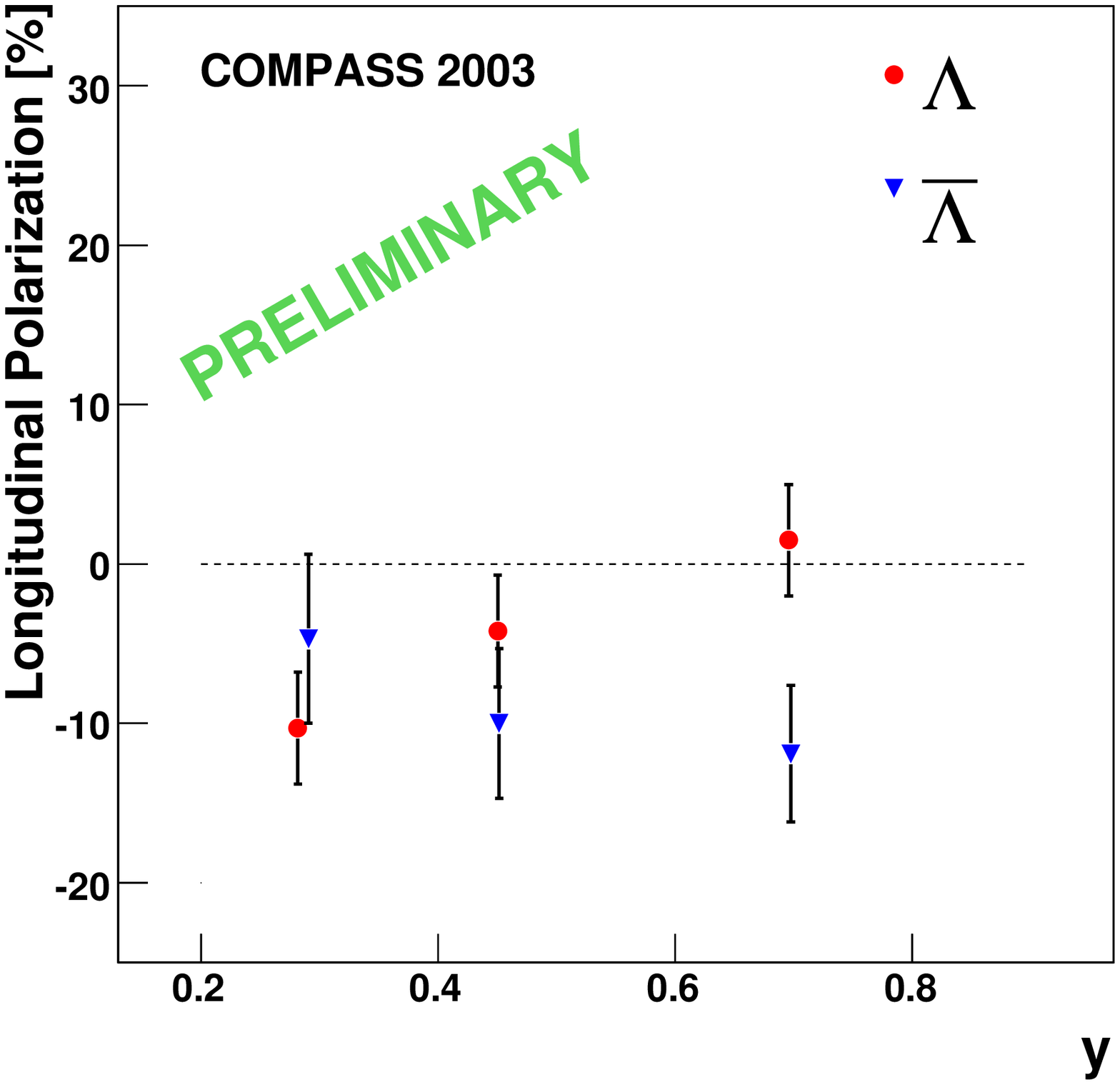,width=8cm,height=9cm}}&
\mbox{\epsfig{figure=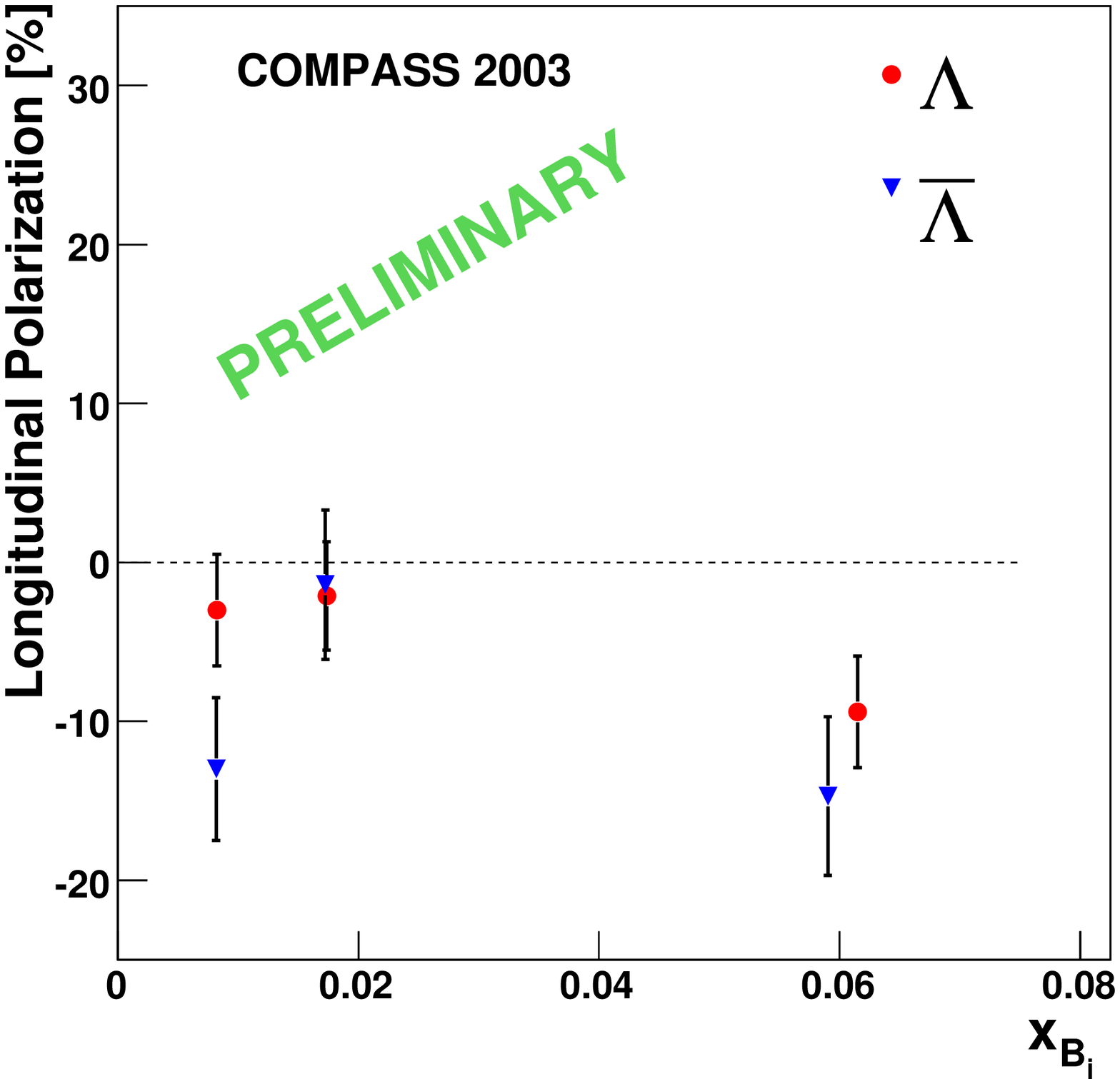,width=8cm,height=9cm}}\\
 {\bf(a)}& {\bf(b)}
\end{tabular}
\end{center}
{\small{\bf Figure 4.} The longitudinal polarization of $\Lambda$
and $\bar{\Lambda}$ for different $y$ (a) and $x$ (b) regions. The
errors are statistical.}
\end{figure}

%\begin{wrapfigure}[21]{R}{5cm
%\begin{center}
%\mbox{\epsfig{figure=sapozh_fig4.eps,width=5cm,height=5cm}}
%\end{center}
%{\small{\bf Figure 4.} The longitudinal polarization of $\Lambda$
%and $\bar{\Lambda}$ for different $y$ regions. The errors are
%statistical. } \label{y}
%\end{wrapfigure}

One could see that, in general, the longitudinal polarizations of
\la~ and \al~are the same. The exception is the region of large $y$,
where there is an indication that the polarization of \la~and \al~
is different. It is interesting that the \la~polarization is large
at small $y$ and decreases with $y$, whereas the \al~polarization
seems to be constant or increasing with $y$. It indicates on
different mechanisms of \la~and \al~ production and polarization.

 The $x$-dependence of the longitudinal polarization is
shown in Fig. 4b. The data sample  has been divided it 3 bins: $x <
0.0121$, $0.0121 \leq x < 0.025$ and $x>0.025 $. The number of
events in each bin is approximately the same. One could see that the
polarizations of \la~ and \al~ are the same except, probably, the
region of small $x<10^{-2}$. Bearing in mind that our $Q^2$ region
is not large and for the fixed $Q^2$ there is a correlation between
$x$ and $y$ variables, it is possible that the difference in \la~and
\al~ polarizations at small $x$ is a reflection of the difference at
large $y$.

The longitudinal polarization of \la~and \al~ hyperons has been
calculated for the COMPASS energy region in
\cite{Kot.98},\cite{Ell.02},\cite{Liang.05}. The main prediction of
\cite{Liang.05}, which follows from the eq.(\ref{plam}), is that the
longitudinal polarizations of \la~and \al~ must be the same if
$q(x)=\bar{q}(x)$ and $\Delta q(x)= \Delta \bar{q}(x)$. In general,
the results of Fig.4 confirm this prediction, though with some
caveats on different $y$-dependence of \la~and \al~polarizations.

In \cite{Ell.02} the longitudinal polarization of $\Lambda$ was
studied on the basis of static SU(6) quark-diquark wave functions
and polarized intrinsic strangeness model
\cite{Ell.95},\cite{Ell.00}. The free parameters of the calculation
are fixed by fitting NOMAD data \cite{NOMAD} on the longitudinal
polarization of \la~ hyperons in neutrino collisions. The small
polarization of \la~ was predicted $P_{\Lambda}=-0.4\%$ for $x_F
> -0.2$ and $0.5<y<0.9$. This prediction is in a nice agreement with
the experimental result shown in Fig. 4a (third bin). An important
advantage of the model \cite{Ell.02} is the consideration of both
quark and diquark fragmentation processes. However, the polarization
of \al~was not considered in \cite{Ell.02}.

In \cite{Kot.98} the longitudinal polarization of $\Lambda$ and
\al~was calculated assuming quark fragmentation only. The set of
cuts ($E_{\mu}=200$ GeV, $P_b=-0.8$, $x_F>0$, $z>0.2$, $Q^2
> 4~ GeV^2$ and $0.5 < y < 0.9$)
 did not exactly coincide with the COMPASS conditions. However the
 trend is interesting. It was predicted for \al~polarization that
 $P_{\bar{\Lambda}}=-15\%$ for NQM-based fragmentation function and
$P_{\bar{\Lambda}}=-5\%$ for the BJ-scheme \cite{BJ}. Our
experimental value for the \al~polarization shown in Fig.4a (third
bin), better agrees with the NQM, though other models are not
excluded at the present statistics and systematics levels.

Comparing the experimental results with the existing theoretical
predictions, one may conclude that the longitudinal polarizations of
\la~ and \al~in DIS at the COMPASS energy seems to be the same, at
least, in the low $y$-region. Nevertheless, the production
mechanisms of \la~ and \al~are different. The \la~ hyperons, even in
the current fragmentation region $x_F>0$, are produced with large
contribution from the diquark fragmentation, which destroys typical
$y$-dependence and leads to a small spin transfer from the polarized
lepton to \la. The polarization of \al~ seems to be more promising
for investigation of the mechanisms of spin transfer from quark to
hyperon.

For more definite conclusions, more precise experimental data and
theoretical calculations are needed. In future, the analysis of the
2004 run data will increase the statistics of at least a factor 2.
At the same time we intend to improve the Monte Carlo description of
the spectrometer to decrease the systematics. The theoretical
calculations of \la~ and \al~longitudinal polarizations at the
COMPASS conditions are highly desirable.

%I am grateful to J.Ellis, A.Kotzinian, Liang Zuo-tang and D.Naumov
%for the fruitful discussions.

\end{document}